\documentclass[onecol]{epl2} 
\def\ket#1{|#1\rangle}
\usepackage{amssymb,amsmath}
\usepackage{multirow}
\usepackage{mathtools}
\newcommand{\tr}{{\rm tr}\,}

\title{Formation probabilities   in quantum critical
chains and Casimir effect}
\author{M.~A.~Rajabpour}
\institute{                    
   Instituto de F\'isica, Universidade Federal Fluminense, Av. Gal. Milton Tavares de Souza s/n, Gragoat\'a, 24210-346, Niter\'oi, RJ, Brazil
}

\abstract{
We find a connection between logarithmic formation probabilities  of two disjoint intervals of quantum critical  spin chains and
the Casimir energy of two aligned needles in two dimensional classical critical systems. Using this connection we provide
a formula for the logarithmic formation probability of two disjoint intervals in generic $1+1$ dimensional
critical systems. The quantity is depenedent on the full structure of the underlying conformal field theory and so useful to find the universality class
of the critical system. 
The connection that we find also  provides a very efficient numerical method to calculate the 
Casimir energy between needles
using quantum critical chains. The agreement between numerical results performed on critical transverse field Ising model and XX chain 
with our
exact results is very good.
We also comment on the mutual R\'enyi information of two disjoint intervals.}

\begin{document}

\maketitle

The subjects of quantum phase transitions and finding the universality class of the critical points have been of great interest
in recent years 
both for theoretical reasons and also because of the recent experimental realization of many low-dimensional quantum systems. In quantum  chain studies
the tradition is to find the critical point, then the central charge of the underlying conformal field theory and finally
critical exponents of the system. The same procedure also works for two dimensional 
classical systems. Since the birth of conformal field theory, different methods were invented to follow the above procedure with
each of them having their own  advantages and disadvantages. For example, one of the early methods to calculate the central charge was the transfer matrix 
technique \cite{CardyAffleck} which, although very efficient in numerical calculations, would not give the central charge without having the sound velocity.
Calculations based on entanglement entropy \cite{centralcharge} does not suffer from this problem, however, they are very difficult to measure in actual
experiments. The recent investigations based on  Shannon mutual information \cite{AR} seem to be promising for experimental investigations
but at the same time they are basis dependent quantities. Similar arguments are also valid in calculating extra structures 
of different universality classes such as the critical exponents. Since diverse methods are useful in different setups it seems 
to be natural to investigate wide range of methods for finding the universality class of the system.

Consider the ground state of a critical spin chain $\ket{\psi_G}=\sum_I a_I\ket{I}$ written in a particular local basis, where $I$ here stands
for particular configuration.
The logarithmic formation probability in this basis is defined as the logarithm of the probability of having particular 
configuration  $p_I=|a_I|^2$ in the domain $D$. When $D$ is a connected domain and the corresponding basis is the $\sigma^z$ basis and the 
desired configuration is a sequence of up (down) spins, this quantity is called emptiness formation probability and has been studied 
in many critical and non-critical spin chains in the last couple of decades \cite{Essler,U1,Abanov-Korepin,Shiroishi,Franchini,Stephan2013,NR2015}. 
The results of many analytical calculations suggest that the emptiness formation probability of critical systems decays
like a Gaussian (exponential) for systems with (without) $U(1)$ symmetry. The exponential
is accompanied with a power-law with an exponent which is universal and as we mention later
dependent on the central charge \cite{Stephan2013}. In other words by calculating the emptiness formation probability one
can gain information about the universality class of the system. Although knowing the central charge of the system helps 
in getting some information regarding the universality class of the system, in principle, it is not enough to determine it uniquely.
To fix the universality class, one needs to also have  information regarding the scaling operators of the system. In this letter, we will
show that  extra information can be achieved by studying the formation probabilities of two disjoint intervals. From
this perspective, our quantity of interest plays the same role as the entanglement entropy. The entanglement entropy of a connected subsystem
fixes the central charge and the same quantity for two disjoint intervals fixes the operator content of the system, 
see \cite{CC2009} and references therein. To calculate the formation probability of two disjoint intervals, we first map
the system to the problem of the Casimir free energy of two aligned needles.  A great deal is known about the form of
the Casimir energy of floating objects in two dimensional medium \cite{Machta,Kardar,needle}, see also \cite{Casimir} and references therein.
Using the results of \cite{Machta,Kardar} we give a formula for the formation probability of two disjoint intervals in 
one dimensional critical quantum systems. We then check our finding using the numerical technique developed in \cite{NR2015}
for the critical Ising model and the XX chain. Our mapping between the formation probabilities and
the Casimir energy of the aligned needles gives a very efficient method to also calculate numerically the Casimir energy of the needles which
is extremely difficult in the usual two dimensional classical setups. For recent connection between the Casimir effect and 
the $n=2$ R\'enyi entanglement entropy  see \cite{Maghrebi}. The techniques based on Casimir effect have also found  many applications
in recent studies of non-equilibrium statistical mechanics \cite{Gambassi}.

In \cite{Stephan2013} by mapping the problem of critical quantum chain to a two dimensional classical counterpart  it was shown
that  the emptiness formation probability of a subregion for a spin chain can be seen as the ratio of two partition functions $\frac{Z^{slit}}{Z}$,
where $Z^{slit}$ is the partition function of the whole system minus a slit and $Z$ is the total partition function. The argument goes as follows:
consider a quantum periodic system with the Hamiltonian $H$. The transfer matrix of the system is  $T=e^{-\epsilon H}$, where $\epsilon$ is 
the imaginary time step. Since in the limit of 
infinite steps $N\to\infty$ we have $T^N\sim e^{-\epsilon NE_g}|\psi_g><\psi_g|$ one can now calculate the emptiness formation probability of the ground
state
using the following formula:
\begin{eqnarray}\label{formation probabilities definition}
 p_{D}=\text{lim}_{_{N\to\infty}} \frac{<\psi|T^{\frac{N}{2}}\delta(|\mathbf{\sigma}>-|\uparrow...\uparrow>_{b})T^{\frac{N}{2}}|\psi>}{<\psi|T^N|\psi>}
\end{eqnarray}
where $\delta(|\mathbf{\sigma}>-|\uparrow...\uparrow>_{b})$ fixes the spins in the basis $b$ in the up direction and $|\psi>$ is the state at infinity
and can be 
in principle any state not orthogonal to
the ground state. The denominator of the above formula is just the total partition of the corresponding two dimensional classical system
on the infinite cylinder. The numerator is dependent on the basis that one chooses and in the most simple case of the
basis which corresponds to classical variables in the corresponding classical system, for example the $\sigma^x$ basis
in the transverse field Ising model, it is just a partition function of a two dimensional classical system
with fixed boundary conditions on a slit. We will later show that if one chooses the $\sigma^z$ basis in the Ising model this
will lead us to free boundary conditions on the slit.
If the boundary condition
imposed by the chosen basis  flows to  some sort of boundary conformal field theory (BCFT) then one can calculate
the logarithmic emptiness formation probability using BCFT techniques. Consider $p_D$ as the probability of formation then we have 
the following formula for logarithmic emptiness formation probability $\Pi_{D}$ of a subsystem with size $s$\cite{Stephan2013} 
\begin{eqnarray}\label{formation probabilities}
\Pi_{D}:=-\ln p_{D}=\alpha s+\frac{c}{8}\ln s+...,
\end{eqnarray}
where $\alpha$ is a non-universal constant, $c$ is the central charge of the system and the dots are the subleading terms.
In \cite{NR2015} it was shown that the method not only works
for the emptiness formation probability but also for much more general formation probabilities. This is just because there
are many different configurations that flow to BCFT in the scaling limit.
It is quite obvious that the method can be generalized 
to cases in which the domain $D$ is not a connected domain. The next simple case is the formation probability of 
two-disjoint intervals $A$ and $B$. In this case, consider  the configurations  $\mathcal{C}_A$ and $\mathcal{C}_B$ in the
intervals $A$ and $B$ respectively. First of all in principle
the configurations $\mathcal{C}_A$ and $\mathcal{C}_B$ not only can be different, they also 
can be considered in different bases. Following \cite{Stephan2013} and the argument given for one interval  one can simply write
\begin{eqnarray}\label{conformal map}
\mathcal{R}(A,B):=-\ln\frac{p(\mathcal{C}_A,\mathcal{C}_B)}{p(\mathcal{C}_A)p(\mathcal{C}_B)}=-\ln\frac{Z^{AB}Z}{Z^AZ^B},
\end{eqnarray}
where $\mathcal{R}$ is defined based on the first equality.  $p(\mathcal{C}_A,\mathcal{C}_B)$ is the corresponding probability of two disjoint intervals $A$ and $B$ and $p(\mathcal{C}_A)$, $p(\mathcal{C}_B)$ 
are the corresponding probabilities of the intervals $A$ and $B$.
$Z^{A}$, $Z^{B}$ and $Z^{AB}$ are the partition functions of the whole $2d$ classical system with slits on $A$, $B$ and $AB$ respectively
and $Z$ is the total partition function. Note that the normalization is in a way that for $A$ and $B$ very far from each other $\mathcal{R}(A,B)$
goes to zero. It is worth mentioning that depending on the configurations  $\mathcal{C}_{A(B)}$ and the bases the partition functions
$Z^{A(B)}$ can be different because different configurations in different bases can induce different boundary conditions on the slits. However, as we
will show in the example of the Ising model there are a lot of configurations that in the scaling limit flow to the same boundary conditions
which means that the leading universal term of $\mathcal{R}(A,B)$ is the same for all of these configurations.

\begin{figure} [htb] \label{fig1}\centering
\includegraphics[width=0.6\textwidth]{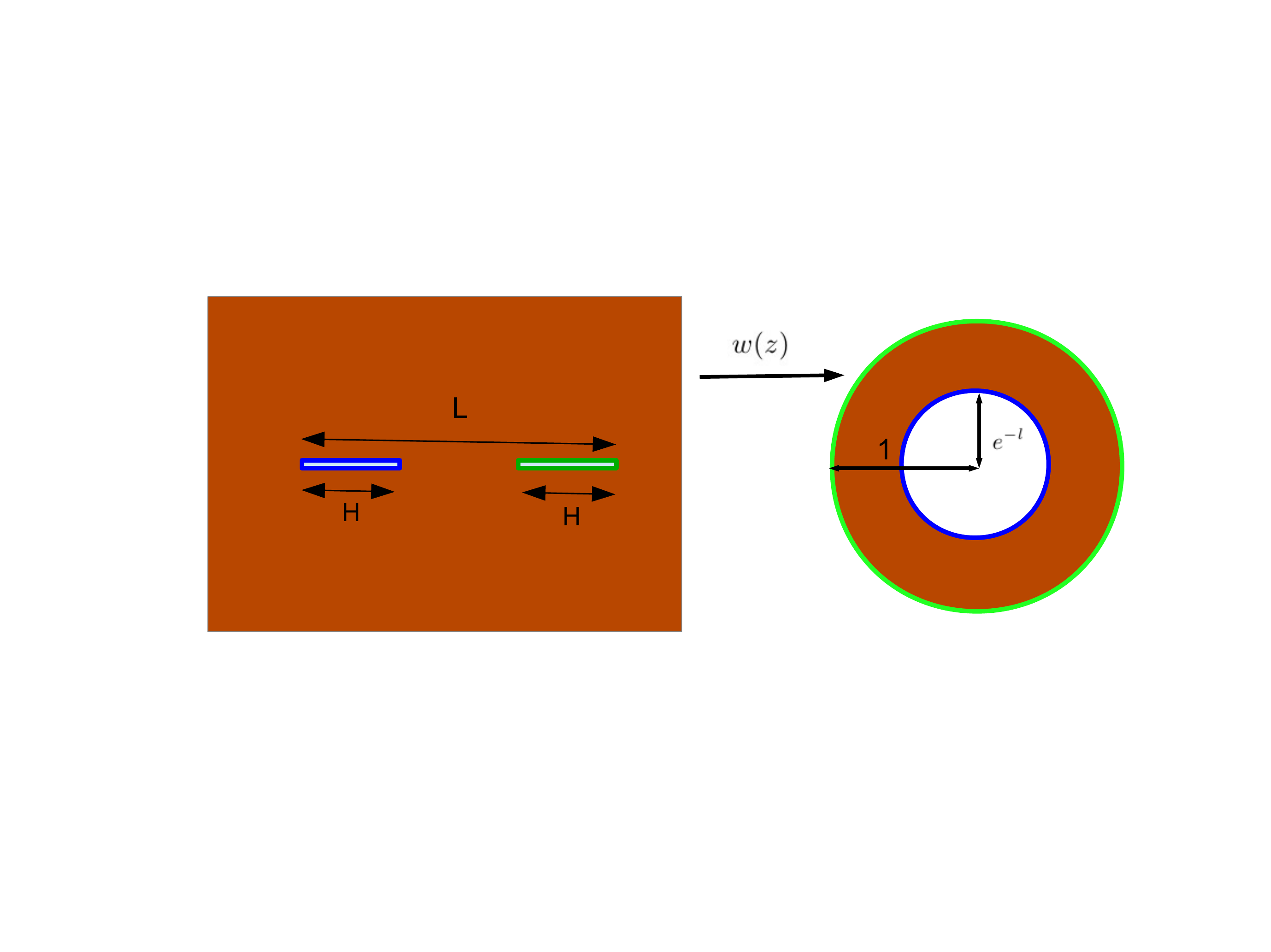}
\caption{The conformal map $w(z)$ sends the plane with two slits to the annulus.} 
\end{figure}
It is not difficult to see that the  quantity $\mathcal{R}(A,B)$, by definition \cite{Machta}, is just the Casimir free energy of two needles in a $2d$ system. The Casimir free energy
for arbitrary shapes in two dimensional critical systems is calculated using conformal field theory techniques in \cite{Machta,Kardar}. The idea is based on
mapping the $2d$ system with two removed domains to the annulus and then using the partition function of CFT on the annulus. Following this procedure
the Casimir free energy can be written as \cite{Machta,Kardar}:
\begin{eqnarray}\label{casimir energy}
\mathcal{R}(A,B)=\mathcal{F}_{ann}+\mathcal{F}_{geo},
\end{eqnarray}
where $\mathcal{F}_{ann}$ is the free energy of CFT on the annulus which is known for most of CFT's \cite{Cardy}
and $\mathcal{F}_{geo}$ can be calculated using \cite{Kardar}
\begin{eqnarray}\label{integral}
\frac{\delta \mathcal{F}_{geo}}{\delta H}=\frac{ic}{6\pi}\oint_{\partial S_2}\{w,z\}dz,
\end{eqnarray}
where $S_2$ is a contour surrounding one of the domains $A(B)$ and $\{w,z\}=\frac{ w'''}{w'}-\frac{3}{2}(\frac{w''}{w'})^2$ is the Schwarzian derivative.
Note that in this work we keep $L$ fixed and vary $H$. 
To calculate the logarithmic formation probability
of two disjoint intervals one just needs to apply the above equations to the problem
of two slits.
The conformal map from the plane with two cuts on a line with lengths $H$ and the distance $L-2H$ to the annulus with the inner 
and outer radiuses $r=e^{-l}$ and $r=1$, see Figure ~1, is given by\footnote{The extension of this result to non-equal  slits is straightforward.} 
\begin{eqnarray}\label{conformal map}
w(z)&=&e^{-\frac{l}{2}}e^{l \frac{\text{sn}^{-1}(\frac{2z}{d},k^2)}{2\mathcal{K}(k^2)}},\\
l&=&2\pi\frac{\mathcal{K}(k^2)}{\mathcal{K}(1-k^2)},
\end{eqnarray}
where $\mathcal{K}$ and $\text{sn}^{-1}$ are
the elliptic  and inverse Jacobi functions \footnote{Note that in all of the formulas we adopt
the Mathematica convention for all the elliptic functions.} respectively and $k=\frac{d}{L}$ and $d=L-2H$.
After performing the integration, we will finally have
%\begin{widetext}
\begin{eqnarray}\label{integral}
\mathcal{F}_{geo}=\frac{c}{12L}\int_a^H dh \frac{L^2\pi^2-2\big{(}h^2-4hL+2L^2\Big{)}\mathcal{K}^2(\frac{4h(L-h)}{L^2})}
{\Big{(}2h^2-3hL+L^2\Big{)}\mathcal{K}^2(\frac{4h(L-h)}{L^2})}
\end{eqnarray}
%\end{widetext}
where $a$ plays the role of lattice spacing.

The annulus contribution part for a generic CFT with boundary conditions $b$ and $d$ on the two boundaries
has the following form \cite{Cardy}
\begin{eqnarray}\label{annulus CFT}
\mathcal{F}_{ann}=-\ln Z_{bd}(\tilde{q})=c\frac{l}{12}-\ln \sum_{h}|b_{h}|^2\chi_h(\tilde{q}^2),
\end{eqnarray}
where $\tilde{q}=e^{-l}$ and $Z_{bd}(\tilde{q})$ is the partition function on the annulus. $\chi_h(\tilde{q}^2)$ is the character of the conformal operator with conformal weight $h$ 
and $b_{h}$ is a number.
It is worth mentioning that for large distances we will have $\mathcal{F}_{ann}\sim \frac{1}{L^{2x_{bd}}}$, where $x_{bd}$
is the smallest eigenvalue of $L_0+\bar{L}_0$ that couples the boundary conditions $b$ and $d$ \cite{Kardar}. $L_0$ and $\bar{L}_0$ 
are the Virasoro generators.
For the Ising model with free-free (FF) boundary conditions one can write the above formula explicitly as follows:
\begin{eqnarray}\label{annulus Ising}
\mathcal{F}_{ann}^{FF}=\frac{l}{24}-\frac{1}{2}\log\frac{\Theta(3,e^{-l})}{\eta(\frac{i l}{\pi})},
\end{eqnarray}
where $\Theta$ and $\eta$ are the Jacobi and Dedekind functions respectively. For the $XX$-chain the free energy on the annulus for the Dirichlet-Dirichlet
(DD) boundary condition
is \cite{Bilstein}
\begin{eqnarray}\label{annulus XX}
\mathcal{F}_{ann}^{DD}=\frac{l}{12}-\log\frac{\sum_{_{n\in
Z}}e^{-n^2l}}{\eta(\frac{i l}{\pi})}.
\end{eqnarray}
We now check the above relations numerically using the method proposed in \cite{NR2015} for the critical XY chain.
The Hamiltonian of XY chain is given by
\begin{eqnarray}\label{Hamiltonian Ising}
H=-\sum_{j=1}^L\Big{[}(\frac{1+\gamma}{2})\sigma_j^x\sigma_{j+1}^x+(\frac{1-\gamma}{2})\sigma_j^y\sigma_{j+1}^y+h\sigma_j^z\Big{]}.
\end{eqnarray}
After the Jordan-Wigner transformation it can be written as
\begin{figure} [htb] \label{fig2}
\centering
\includegraphics[width=0.6\textwidth]{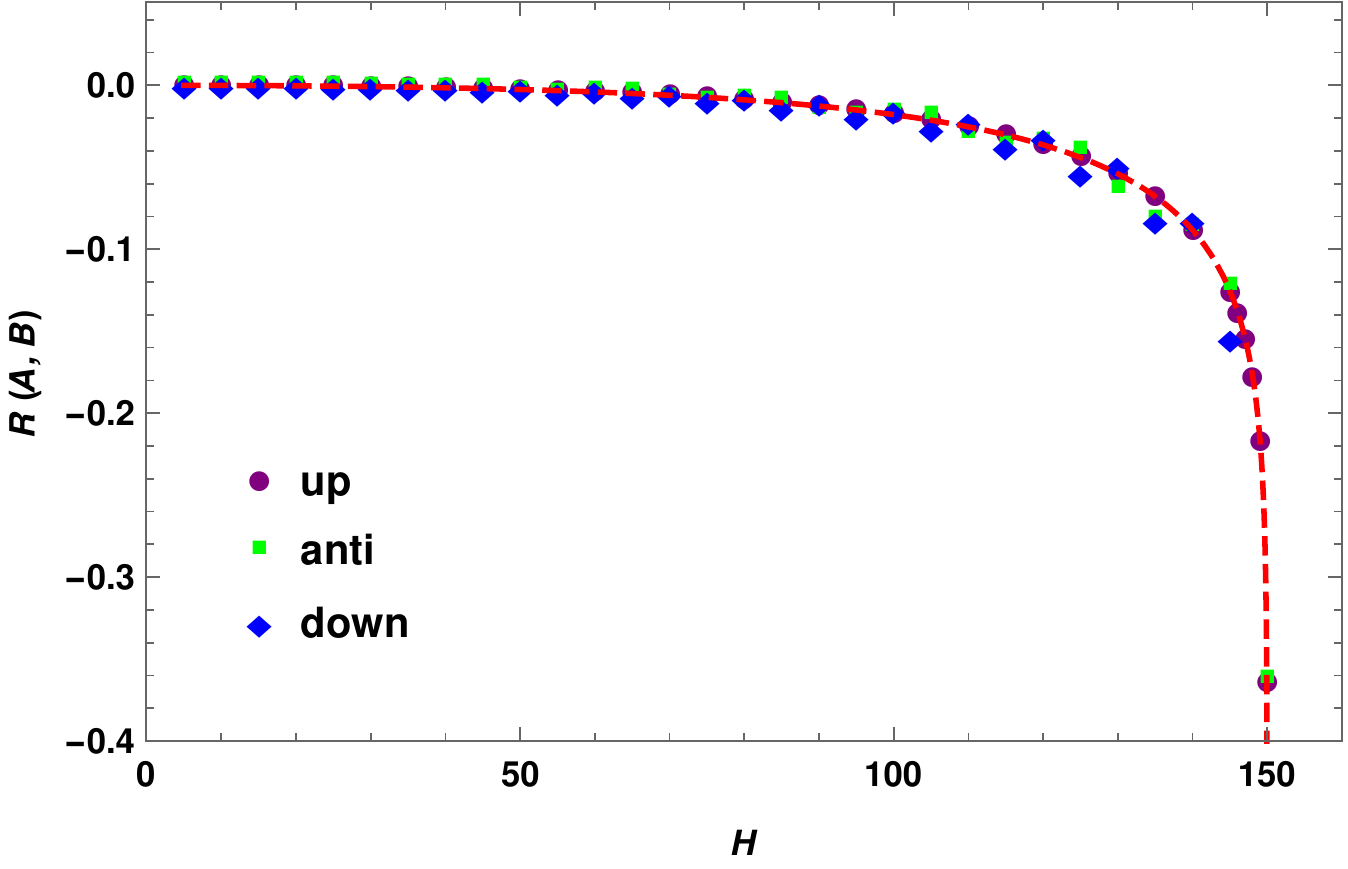}
\caption{ The quantity $\mathcal{R}$ for two disjoint intervals for configurations with, all spins up, all spins down and antiferromagnetic pattern.
 The dashed line is the analytical result. The parameter $L=300$ is fixed in our calculations.} 
\end{figure}
\begin{equation}\label{Ising Hamiltonian free fermion}
H=-\sum_{j=1}^L \Big{[}(c_j^{\dagger}c_{j+1}+\gamma c_j^{\dagger}c_{j+1}^{\dagger}+h.c.)+h(2c_j^{\dagger}c_j-1)\Big{]},
\end{equation}
where for $\gamma=h=1$ we have the critical Ising model and for $\gamma=h=0$ we have the critical XX chain. Note that in the language
of free fermions a spin up  (down) in the $\sigma^z$ basis means the lack (presence) of a fermion in that particular site.
Using the coherent states it was shown in \cite{NR2015} that the probability of configuration $\mathcal{C}$
for an arbitrary region can be calculated using the following formula
\begin{eqnarray}\label{Formation probability}
P(\mathcal{C})=\det[\frac{1}{2}(1-G)] M_J^{\mathcal{C}},
\end{eqnarray}
where the block Green matrix is defined as $G_{ij}=\tr[\rho_Db_ia_j]$ with $a_i=c_i^{\dagger}+c_i$ and $b_i=c_i^{\dagger}-c_i$.
$M_J^{\mathcal{C}}$ is the minor of the matrix $J=(G+1)(G-1)^{-1}$ corresponding to the configuration $\mathcal{C}$. Different
configurations correspond to different diagonal minors of the matrix $J$, for example, for the configuration of all  spins up the emptiness formation
probability will be just $\det[\frac{1}{2}(1-G)] $ which we used the minor with the rank $k=0$. 
The method to extract the probability of other configurations are explained in \cite{NR2015}. The algorithm of the calculation is as follows:
for any spin up in the configuration one just needs to remove the corresponding row and column in the matrix $J$ and then calculate the determinant.
We applied this procedure to calculate the formation probability of two disjoint intervals in the critical Ising model with
$G_{ij}=-\frac{1}{\pi(i-j+1/2)}$. It is expected that
the configurations of all spins up (down) and alternating antiferromagnetic configuration flow to free-free conformal boundary conditions in
the scaling limit. In the case of all spins up the argument for one interval goes as follows: consider
an interval with all spins up in the $\sigma^z$ basis. To make contact with the classical Ising model we need to write the state with all
the spins up in the $\sigma^x$ basis as follows:
\begin{eqnarray}\label{Free BC}
|\uparrow...\uparrow>_{z}=\frac{1}{2^{s/2}}(|\rightarrow>_x+|\rightarrow>_x)...(|\rightarrow>_x+|\rightarrow>_x)=
\frac{1}{2^{s/2}}\sum_{\{\sigma_j^x\}}|\sigma_1^x...\sigma_s^x>=|\text{free}>
\end{eqnarray}
The above formula demonstrate that all the spins up configuration  in the Euclidean version is just a free boundary condition.
Generalization of the above argument to arbitrary intervals is straightforward.
Although the above calculation is exact for the case of all spins up the same argument  does not go
smoothly for all spins down and antiferromagnetic configurations. By numerical calculations it is shown in \cite{NR2015} that these configurations
also flow to conformal boundary conditions. Here we show that numerical calculations are consistent with the free boundary conditions 
for all the considered configurations in the $\sigma^z$ basis. Our numerical results depicted in the Figure ~2 show that  the $\mathcal{R}(A,B)$ is
given by (\ref{casimir energy}) 
with $\mathcal{F}_{geo}$ and $\mathcal{F}_{ann}$ given by (\ref{integral}) and (\ref{annulus Ising}). We also realized that the contribution 
coming from $\mathcal{F}_{geo}$ (which we evaluated numerically) is extremely small and one can get  compatible results just by 
considering the $\mathcal{F}_{ann}$. However, one should notice that this term can be significant for needles in very small distances. It is worth
mentioning that the two configurations - all the spins down and the antiferromagnetic one - show  small oscillating behaviour which is a consequence
of the parity effect \cite{NR2015}.
\begin{figure} [htb] \label{fig3}
\centering
\includegraphics[width=0.6\textwidth]{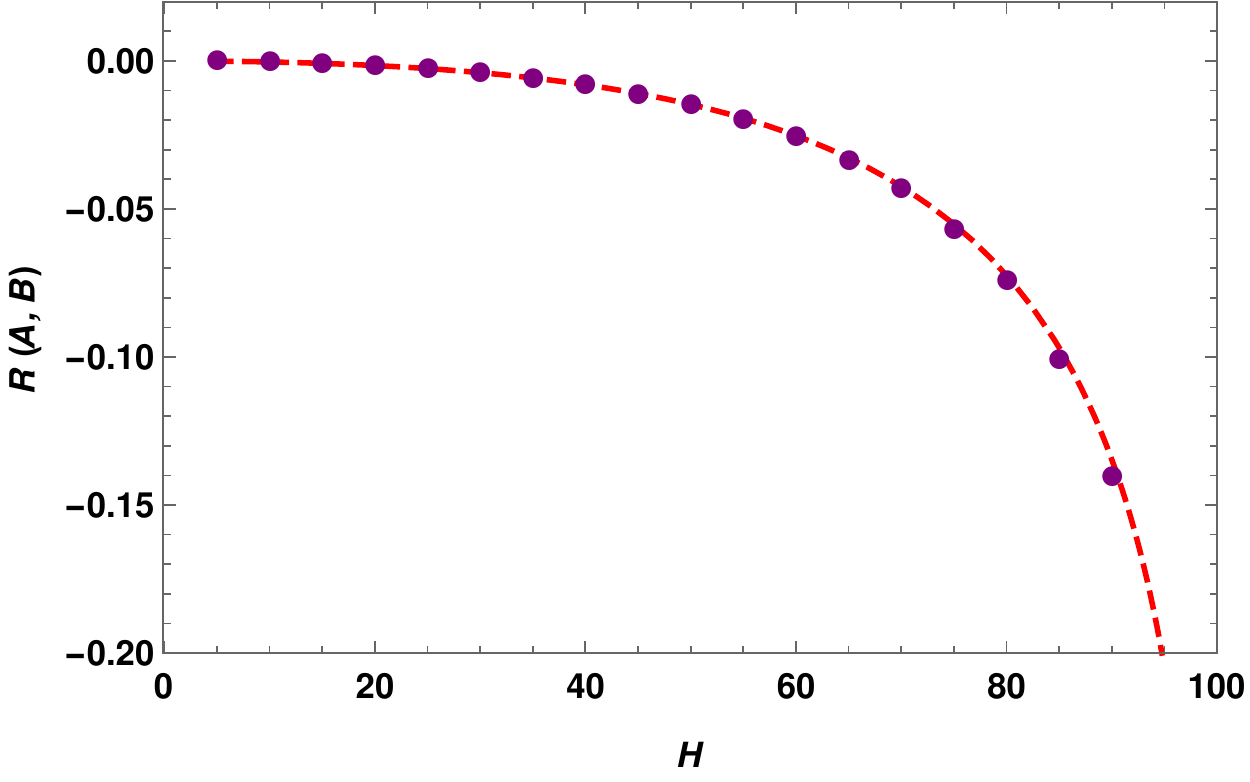}
\caption{The quantity $\mathcal{R}$ for  two disjoint intervals in the XX chain with respect to $H$ for antiferromagnetic configuration. The dots are the data coming from
numerical calculations and the solid line is the analytical result. The parameter $L=200$ is fixed in our calculations.} 
\end{figure}
In the case of XX chain, we considered just the alternating configuration which is expected to flow to the DD boundary condition
in the Luttinger liquid language \cite{Bilstein}. Note that the configuration with all the spins up does not flow to BCFT and so it is
not going to follow the above formulas. The reason behind this fact is that XX chain has a $U(1)$ symmetry which means
that the number of fermions is fixed.
To have a configuration with all spins up  one needs to have a string of empty sites which is not compatible with the half filling picture of XX chain, 
for more precise argument based on arctic circle phenomena see \cite{Stephan2013}. Note that the antiferromagnetic configuration is
perfectly consistent with the $U(1)$ symmetry of the system.
We repeated the same procedure as above  with $G_{ij}=(1-\delta_{ij})\frac{2}{\pi(i-j)}\sin\frac{\pi(i-j)}{2}$ for antiferromagnetic
configuration. The results
depicted in Figure ~3 show that the $\mathcal{R}(A,B)$ is given by (\ref{casimir energy}) 
with $\mathcal{F}_{geo}$ and $\mathcal{F}_{ann}$ given by (\ref{integral}) and (\ref{annulus XX}).

We finally comment on the R\'enyi mutual information of two disjoint intervals. The R\'enyi entropy of the region ${\cal X}$ is defined as 
\begin{equation} \label{Renyi1}
Sh_n({\cal X}) =\frac{1}{1-n}\ln \sum_I p_I^n,
\end{equation}
where $p_I$ is the probability of having a particular configuration in the system and the sum is over all the possibilities.
Then the naive definition of the 
R\'enyi mutual information between the regions $A$ and $B$ is as follows\footnote{For the distinction between different mutual R\'enyi informations see \cite{AR2015} }: 
\begin{equation} \label{Renyi2}
I_n(A,B)= Sh_n(A)+Sh_n(B)-Sh_n(A\cup B).
\end{equation}
As it is argued in \cite{Stephan2013} when $n\to\infty$ only the biggest contribution survives and so one can drop the sum in the definition
of the R\'enyi information. The configuration with the highest probability in the case of the transverse-field Ising model 
is the one with all the spins up. However, For XX-chain the most important probability is 
the antiferromagnetic configuration. As we already discussed both cases are related to BCFT and their formation probabilities
are given by the formulas that we discussed above.
After a bit of algebra, for $n\to\infty$ we have
\begin{equation} \label{Renyi2}
I_n(A,B)= \frac{n}{1-n}\mathcal{R}(A,B).
\end{equation}
Although the above result is derived for $n\to\infty$,  its validity, as in the case of one interval \cite{Stephan2014,AR}, might extend up to finite $n>1$.
%It will be  interesting to check numerically to what extend the above formula works for finite $n$.

\textit{Conclusions:} 
The simple connection  between the formation probabilities and the Casimir effect that we presented here enabled us to find an
exact solution for the formation probabilities of two disjoint intervals in  generic one dimensional quantum critical systems. The quantity
is of great interest because it has a lot of information regarding the universality class of the system.  The connection that we presented
also
gives a very efficient method to numerically calculate the Casimir energy between two needles in a two dimensional medium. 
We also commented on the connection between R\'enyi mutual information and the Casimir effect.
It will be very 
interesting to generalize the results to different models with different boundary conditions. For example studying the Neumann-Neumann
boundary condition in the XX chain \cite{Bilstein} is of great interest. In addition, here we just commented on the formation probabilities of
 the XY-chain in the $\sigma^z$ basis, one needs to also study other bases such as $\sigma^x$ basis which in the case of the Ising model
 is just the fixed-fixed boundary condition \cite{Machta}. Numerical
 study of other bases is a big challenge because one cannot in general use the free fermion representation to calculate them. 
 Finally it will be also interesting to extend some of these results
to higher dimensions.

\acknowledgments
I thank M F Maghrebi, K Najafi and T Oliveira for discussions. This work  was supported in part by
CNPq.

\end{document}